\documentclass[sigconf]{acmart}

\usepackage{makecell}
\usepackage[many]{tcolorbox}
\usepackage{multicol}

\usepackage{adjustbox}

\usepackage{enumitem}

\usepackage{comment}
\usepackage{balance}

\usepackage{fontawesome}
\definecolor{light-gray}{gray}{0.95}

\AtBeginDocument{%
  \providecommand\BibTeX{{%
    \normalfont B\kern-0.5em{\scshape i\kern-0.25em b}\kern-0.8em\TeX}}}

\copyrightyear{2022}
\acmYear{2022}
\setcopyright{acmlicensed}\acmConference[FAccT '22]{2022 ACM Conference on Fairness, Accountability, and Transparency}{June 21--24, 2022}{Seoul, Republic of Korea}
\acmBooktitle{2022 ACM Conference on Fairness, Accountability, and Transparency (FAccT '22), June 21--24, 2022, Seoul, Republic of Korea}
\acmPrice{15.00}
\acmDOI{10.1145/3531146.3533218}
\acmISBN{978-1-4503-9352-2/22/06}

\begin{document}\sloppy

\title[Perceptions of Informational Fairness and Trustworthiness in Automated Decision-Making]{``There Is Not Enough Information'': On the Effects of Explanations on Perceptions of Informational Fairness and Trustworthiness in 
Automated Decision-Making}

\author{Jakob Schoeffer}
\affiliation{
    \institution{Karlsruhe Institute of Technology}
    \country{Germany}
}
\email{jakob.schoeffer@kit.edu}
\orcid{0000-0003-3705-7126}

\author{Niklas Kuehl}
\affiliation{%
  \institution{Karlsruhe Institute of Technology}
  \country{Germany}
}
\email{niklas.kuehl@kit.edu}
\orcid{0000-0001-6750-0876}

\author{Yvette Machowski}
\affiliation{%
  \institution{Karlsruhe Institute of Technology}
  \country{Germany}}
\email{yvette.machowski@alumni.kit.edu}
\orcid{0000-0002-9271-6342}


\begin{abstract}
Automated decision systems (ADS) are increasingly used for consequential decision-making.
These systems often rely on sophisticated yet opaque machine learning models, which do not allow for understanding how a given decision was arrived at.
In this work, we conduct a human subject study to assess people's perceptions of \emph{informational fairness} (i.e., whether people think they are given adequate information on and explanation of the process and its outcomes) and \emph{trustworthiness} of an underlying ADS when provided with varying types of information about the system.
More specifically, we instantiate an ADS in the area of automated loan approval and generate different explanations that are commonly used in the literature.
We randomize the amount of information that study participants get to see by providing certain groups of people with the same explanations as others \emph{plus} additional explanations.
From our quantitative analyses, we observe that different amounts of information as well as people's (self-assessed) AI literacy significantly influence the perceived informational fairness, which, in turn, positively relates to perceived trustworthiness of the ADS.
A comprehensive analysis of qualitative feedback sheds light on people's desiderata for explanations, among which are (i) consistency (both with people's expectations and across different explanations), (ii) disclosure of monotonic relationships between features and outcome, and (iii) actionability of recommendations.
\end{abstract}



\maketitle

\section{Introduction}
Automated decision-making has become ubiquitous in many high-stakes domains such as hiring \cite{kuncel2014hiring}, bank lending \cite{townson2020ai}, grading \cite{satariano2020british}, and policing \cite{heaven2020predictive}, among others.
The underlying motives of adopting automated decision systems (ADS)\footnote{A summary of our abbreviations is given in Tab.~\ref{tab:abbrev} in \S~\ref{sec:abbrev}} are manifold: they range from cost-cutting to improving performance and enabling more robust and objective decisions \citep{harris2005automated,kuncel2014hiring,Newell2015}.
Hopes are also that, if properly designed, ADS can be a valuable tool for breaking out of vicious patterns of human stereotyping and contributing to social equity, e.g., in the realms of recruitment \citep{chalfin2016productivity,koivunen2019understanding}, health care \citep{grote2020ethics,triberti2020third}, or financial inclusion \citep{lepri2017tyranny}.
However, ADS are typically based on ML techniques, which, in turn, rely on historical data.
If, e.g., this underlying data is biased (e.g., because certain socio-demographic groups were favored in a disproportionate way), an ADS will learn from and perpetuate existing patterns of unfairness \citep{feuerriegel2020fair}.
Prominent examples of such behavior from the recent past are race and gender stereotyping in job ad delivery \citep{Imana21a}, as well as the discrimination of Latinx and African-American borrowers in algorithmic mortgage loan pricing \citep{bartlett2021consumer}.
These and other cases have put ADS under enhanced scrutiny, justifiably jeopardizing trust in these systems \cite{Edelman2021}.

In recent years, a growing body of AI and ML research has been devoted to detecting, quantifying, and mitigating unfairness in ADS \cite{mehrabi2019survey}.
A significant share of this work has focused on formalizing different concepts of \emph{fairness} through statistical equity constraints, many of which are at odds with each other~\cite{kleinberg2016inherent,chouldechova2017fair}.
As a consequence, there cannot be a one-size-fits-all technical fairness criterion.
Moreover, in many cases, these techno-centric works do not explicitly take into account the opinions of people that are (potentially) affected by such automated decisions.
While the FAccT community has made a plethora of impactful contributions over the past years, it is still crucial to better understand people's perceptions and attitudes towards ADS---in addition to how researchers may define those systems' fairness in technical terms.

A related issue revolves around explaining automated decisions to affected individuals.
As ADS employ ever more sophisticated and ``black-box'' ML models, several problems arise; one of which is the hampered detectability of adverse behavior of such systems.
In order to safeguard transparency and accountability of automated decisions, several laws and regulations demand a ``right to explanation''.
The EU General Data Protection Regulation (GDPR), e.g., requires the disclosure of ``the existence of automated decision-making, including [\dots] meaningful information about the logic involved [\dots]'' \cite{GDPR} to data subjects.
In fact, it has been shown, among others, that explanations can enhance people's understanding of certain automated decisions \citep{lim2009and}.
For most real-world cases, however, those regulations generally remain (too) vague and little actionable---which often results in deficient adoption, as noticed in the context of bank lending \cite{poland}.
Moreover, research on \emph{explainable AI} (XAI) suggests that there exists no one-size-fits-all approach to explaining ADS either \citep{langer2021we,arya2019one}.

In this work, we conduct a human subject study to examine the effects of explanations on people's perceptions towards an automated loan approval system, where we randomize the type and amount of information that study participants get to see.
The primary dependent variables that we are interested in are perceptions of \emph{informational fairness} of the system (i.e., whether people think they are given adequate information on and explanation of the decision-making process and its outcomes) as well as perceived \emph{trustworthiness}, and the relationship between both.
We also assess the influence of people's (self-assessed) \emph{AI literacy} on the outcomes.
Finally, we ask multiple open-ended questions w.r.t. people's ability to assess the given system's fairness, as well as regarding the appropriateness of explanations' content.

\section{Background and Related Work}
Topics of \emph{fairness} and \emph{trustworthiness} have become important pillars of AI and HCI research in recent years.
In this section, we provide an overview of relevant literature and highlight our contributions.
For brevity, we do not explicitly cover the vast technical literature on algorithmic fairness.
While we assume that the FAccT community is familiar with seminal work in this field, we refer interested readers from other disciplines to relevant survey literature: \cite{mehrabi2019survey,barocas2018fairness,caton2020fairness}.

It is---albeit unsurprisingly---important to note that a ``fair'' (according to some technical fairness notion) system does not imply that people perceive it as such; either because their personal fairness concepts differ from the employed technical notion or because they are not enabled to assess the system's (un)fairness, to begin with.
In fact, it must be questioned whether an ADS that satisfies given statistical notions of fairness (e.g., equitable distribution of outcomes) can ever be \emph{truly} considered fair when at the same time decision-subjects are left in the dark w.r.t. the inner workings of the system.
Instead, \emph{fairness} (of ADS) is likely a multi-faceted construct that encompasses different dimensions, similar to dimensions of (organizational) justice \cite{colquitt2001justice,colquitt2015measuring}, which are commonly made up of distributive, procedural, interpersonal, and informational justice \cite{colquitt2015measuring}.
While distributive and procedural aspects have been considered in the context of ADS (e.g., in \cite{long2021fairness,grgic2018beyond,lee2019procedural}), work on \emph{informational fairness} of ADS is lacking.

Borrowing from \cite{chan2011perceptions}, we call a system \emph{informationally fair} if it conveys adequate information on and explanation of the decision-making process and its outcomes; and we define \emph{adequate information} (similar to \cite{colquitt2015measuring}) as information being \emph{thorough, reasonable, tailored} (to individual needs), as well as helping people \emph{understand} the decision-making process, and enabling them to judge whether this process is fair or unfair.
We refer to  \S~\ref{appendix:constructs_items} for an overview of our measurement items.
\emph{Trustworthiness} is a well-established construct that, according to \cite{belanger2002trustworthiness}, is defined as ``the perception of confidence in the [\dots] reliability and integrity [of an ADS].''
We refer the reader to \cite{lee2004trust,vereschak2021evaluate,jacovi2021formalizing} for survey literature on trust and trustworthiness.

\subsection{Related work}

\paragraph{Automated decision systems}
\citet{harris2005automated} define \emph{automated decision systems} (ADS) as systems that aim to minimize human involvement in decision-making processes.
In this work, we assume ADS to be supervised ML models.
In many cases, ADS have the potential to make more consistent decisions than humans.
Such systems are popular in many industries, such as banking \cite{harris2005automated,townson2020ai} or hiring \cite{carey2016companies,chalfin2016productivity,koivunen2019understanding,kuncel2014hiring}---and they are emerging in new areas as well, e.g., in health care \cite{grote2020ethics,triberti2020third}.
With their increasing adoption in different consequential areas, it is important to ensure that ADS reach fair decisions that are transparent, primarily, to affected individuals or auditors.
However, there have been multiple cases in the recent past where algorithms made biased decisions that discriminated against certain groups, e.g., based on gender or race \cite{angwin2016machine,buolamwini2018gender,heaven2020predictive}.
In other instances, ADS have been operating in an opaque (``black-box'') fashion, making it, among others, difficult (i) for affected individuals to grasp the rationale behind certain decisions, and (ii) for regulatory agencies and other responsible stakeholders to vet such systems appropriately \cite{pasquale2015black}.
On that account, fairness and transparency of ADS have become important topics of interest for the research community. 
Interestingly, despite known weaknesses of ADS, some prior work has found that human-made decisions are \emph{not} generally perceived as fairer or more trustworthy than automated decisions; primarily for reasons of (alleged) consistency in automated decision-making \cite{schlicker2021expect,schoeffer2021perceptions}.

\paragraph{Explainable AI}
Despite being a popular topic of current research, XAI is a natural consequence of designing ADS and, as such, has been around at least since the 1980s \cite{lewis1982role}.
Its importance, however, keeps rising as increasingly sophisticated (and opaque) AI techniques are used to inform ever more consequential decisions.
XAI is not only required by law (e.g., GDPR, ECOA); \citet{eslami2019user}, e.g., have shown that users’ attitudes towards algorithms change when transparency is increased.
In general, both quantity and quality of explanations matter: \citet{kulesza2013too} explored the effects of soundness and completeness of explanations on end users’ mental models and suggest, among others, that oversimplification is problematic.
Recent findings from \citet{langer2021spare}, on the other hand, suggest that in the case of automated job interviews it might make sense to withhold certain pieces of information from applicants in order to not evoke negative reactions.

Even in the presence of explanations, people sometimes rely too heavily on system suggestions \cite{bussone2015role}, a phenomenon commonly referred to as \emph{automation bias} \cite{de2020case,Goddard2014}.
\citet{ehsan2021explainability} have also used the term ``explainability pitfalls'' for any such unanticipated negative effects of explanations (e.g., unwarranted trust \cite{schlicker2021towards}).
Eventually, \citet{chromik2019dark} (inspired by seminal work related to UX design \cite{gray2018dark}) warn that explanations can be exploited to purposefully deceive users for the benefit of \emph{other} stakeholders.
Hence, explanations are by no means the ``silver bullet'' when it comes to solving problems of opaque AI systems \cite{bauer2021expl}.
A comprehensive overview of XAI stakeholders and their distinct desiderata is given by \citet{langer2021we}.
For instance, people affected by automated decisions may be particularly interested in explanations that enable them to evaluate the fairness and trustworthiness of the underlying systems \cite{schoeffer2021appropriate,langer2021we}.
This desideratum is closely linked to informational fairness of ADS \cite{colquitt2001justice}, as introduced earlier.
We refer the interested reader to, among others, \cite{arya2019one,guidotti2018survey,arrieta2020explainable, adadi2018peeking,goebel2018explainable,molnar2020interpretable,langer2021we,miller2019explanation} for more in-depth literature on different XAI techniques and their inner workings.
Regarding the effectiveness of explanations, generally speaking, prior research has primarily focused on comparing individual explanation styles head-to-head (e.g., \cite{binns2018s,dodge2019explaining}), while little work has been done on evaluating the interplay of different styles, including potential complementarity.
\citet{langer2021we} emphasize the sparsity of empirical work w.r.t. the effectiveness of explanations overall.

\paragraph{Perceptions towards ADS}

A relatively new line of research in AI and HCI has started focusing on perceptions of fairness and trustworthiness in automated decision-making.
For instance, \citet{binns2018s} and \citet{dodge2019explaining} compare fairness perceptions in ADS for distinct explanation styles.
Their works suggest differences in effectiveness of individual explanation styles---however, they also note that there does not seem to be a single best approach to explaining automated decisions.
A different line of research has examined people's moral judgments w.r.t. the use of specific features in ADS \cite{grgic2018human,grgic2018beyond}, also with mixed empirical findings.
\citet{lee2018understanding} compares perceptions of fairness and trustworthiness depending on whether the decision maker is a person or an algorithm in the context of managerial decisions.
Their findings suggest that, among others, people perceive automated decisions as less fair and trustworthy for tasks that require typical human skills.
\citet{lee2017algorithmic} explore how algorithmic decisions are perceived in comparison to group-made decisions.
\citet{wang2020factors} combine a number of manipulations, such as favorable and unfavorable outcomes, to gain an overview of fairness perceptions.
An interesting finding by \citet{lee2019procedural} suggests that fairness perceptions decline for some people when gaining an understanding of an algorithm if their personal fairness concepts differ from those of the algorithm.
\citet{woodruff2018qualitative} conducted workshops with people from traditionally marginalized backgrounds, inferring that awareness of unfairness in ADS can substantially affect trust in companies or products.

Some work has also assessed the impact of people's demographics (including gender \cite{pierson2017demographics}), as well as political views and task experience \cite{grgic2020dimensions} on their perceptions.
\citet{saxena2019fairness} examined lay people's perceptions of different technical fairness notions for ADS, suggesting that people prefer notions related to meritocratic fairness \cite{liu2017calibrated,joseph2016fairness}.
Regarding trustworthiness, \citet{kizilcec2016much}, e.g., concludes that it is important to provide the right amount of transparency for optimal trust effects, as both too much and too little transparency can have undesirable effects.
\citet{kastner2021relation} also examined the relationship between explainability and trust(worthiness), urging system designers to engineer for trustworthiness (as opposed to trust), and indicating that explanations can be a crucial toolbox towards that goal.
Regarding perceptions of different social groups, \citet{lee2021included} point out that prior studies have mostly recruited respondents from \texttt{Amazon Mechanical Turk} \cite{paolacci2010running}, which has predominantly white participants \cite{hitlin2016research}---because of this, among other reasons \cite{prolificvsmturk} we have recruited our study participants through \texttt{Prolific}\footnote{\texttt{Prolific} is a crowdworking platform for online research: \url{https://www.prolific.co/}} \cite{palan2018prolific}.

\subsection{Research gaps and our contributions}
We aim to complement prior work to better understand how much of which information should be provided so that people are optimally enabled to understand the inner workings and appropriately assess the fairness and trustworthiness of ADS.
To that end, we conducted a randomized experiment to examine people's perceptions of informational fairness and trustworthiness towards an automated loan approval system, given different combinations of common explanations (relevant factors, factor importance, and counterfactual explanations).
While there exists prior work on trustworthiness perceptions for \emph{individual} explanation styles, we see a significant gap w.r.t. assessing \emph{combinations} of different explanations.
We argue that this is an important gap to fill because different explanations convey different information and will likely have to be leveraged complementarily (i.e., \emph{not} in isolation) in practice.
On a related note, we also set about examining the marginal effects of providing certain explanations \emph{on top of} others---which, to the best of our knowledge, has not been analyzed in depth before.
As a consequence, we alter the \emph{amount of information} that different groups of people get to see.
We do by no means claim to examine these aspects exhaustively, but we hope that our work will be a stepping stone for further research.

Finally, and perhaps most importantly, we shift focus from examining distributive and procedural fairness perceptions to \emph{informational fairness}.
In other words, we do \emph{not} ask people whether they find particular ADS outcomes or procedures fair or not, but---broadly speaking---whether they feel they received sufficient information to \emph{assess} a given system.
This is an important distinction.
Only very few works have considered the informational fairness dimension when experimentally evaluating effectiveness of ADS explanations: \citet{binns2018s} only measure the understandability aspect of informational fairness for individual explanation styles; \citet{schlicker2021expect} and \citet{schoeffer2021perceptions} assess informational fairness perceptions, but with a focus on comparing human with automated decision makers.
\citet{uhde2020fairness} and \citet{brown2019toward} conducted interviews \cite{uhde2020fairness} and workshops \cite{brown2019toward} to infer qualitative statements related to informational fairness; whereby \citet{brown2019toward} explicitly state that ``more research is needed to understand how different elements of algorithmic systems affect perceptions of [\dots] informational justice.''
Empirical work on the interplay of informational fairness and trustworthiness perceptions for ADS is, to our knowledge, entirely novel.
Finally, we also analyze the relationship between study participants' (self-assessed) AI literacy and their perceptions, and we qualitatively examine their answers to open-ended question regarding (in)appropriateness of explanations as well as what information they feel is missing (if any) to properly vet the given ADS.

\section{Research Hypotheses}
The conditions of our experiment comprise different amounts of information that study participants get to see w.r.t. an ADS in the realm of automated loan decisioning.
Regarding the potential effects of varying amounts of information on our dependent variables of perceived informational fairness and trustworthiness, we formulate two research hypotheses based on preliminary qualitative insights w.r.t. people's desire for transparency and information \cite{uhde2020fairness,brown2019toward} as well as prior findings from the psychology literature \cite{lind1983decision,thibaut1975procedural,colquitt2015measuring,colquitt2011justice,houlden1978preference,van1998we}.
First, assuming that explanations are not entirely lacking in content, we conjecture (similar to \cite{uhde2020fairness,brown2019toward}) that more provided information leads to higher informational fairness perceptions.
Regarding effects on trustworthiness perceptions, we note that several factors contribute to a system's fairness \cite{lee2019procedural,colquitt2015measuring}; among these are \emph{consistency} (of decision-making procedures) as well as \emph{process and outcome control} on behalf of decision-subjects \cite{lee2019procedural,dietvorst2018overcoming}.
\emph{Process control} means that decision-subjects have the ``ability to influence what [\dots] data is considered by the decision maker'' \cite{lee2019procedural}, and \emph{outcome control}, borrowing from \cite{houlden1978preference}, refers ``to the ability to appeal or modify the outcome [\dots] once it has been made'' \cite{lee2019procedural}.
While we do not anticipate our employed explanations to readily increase perceptions of outcome control, we conjecture that certain information may enhance assumed process control, which, in turn, affects procedural fairness perceptions \cite{colquitt2015measuring,lee2019procedural} and, ultimately, trust \cite{van1998we}.
\begin{itemize}
    \item[\textbf{H1}] As the amount of information provided increases, perceptions of informational fairness towards the ADS increase.
    \item[\textbf{H2}] As the amount of information provided increases, perceptions of trustworthiness towards the ADS increase.
\end{itemize}
While investigating these relationships, we are not only interested in the effects of our conditions on informational fairness and trustworthiness but also in the relationship between the latter two.
Some prior work has examined the relationship between informational fairness/justice and trust/trustworthiness (e.g., \cite{colquitt2011justice,lance2010organizational,zhu2012service}) in other contexts.
\citet{lance2010organizational} identified a significant positive effect of informational justice on different facets of trustworthiness perceptions in one of their two examined settings in the realm of organizational justice.
Similarly, \citet{zhu2012service}, in the context of customer satisfaction in internet banking, found that informational fairness (as a component of overall systemic fairness) has a positive effect on trust.
Finally, \citet{colquitt2011justice} affirm that ``conventional wisdom on the justice-trust connection'' implies a causal path from (informational) justice to trust, and not the other way round.
While these works address different use cases, we conjecture a positive relationship between informational fairness and trustworthiness perceptions for our ADS setting as well:
\begin{itemize}
    \item[\textbf{H3}] Perceptions of informational fairness relate positively to perceptions of trustworthiness.
\end{itemize}
Experts may have a different attitude towards procedures or phenomena that touch on their area of expertise than non-experts.
Slovic et al. \cite{slovic1987perception,slovic1981perceived}, e.g., found differences in risk perceptions between experts and lay people.
Regarding innovative (food) technologies, \citet{siegrist2008factors} notes that lay people may neither be able to assess risks nor benefits appropriately.
For the specific case of ADS, \citet{wang2020factors} found a significant effect of computer literacy on a mix of procedural and distributive fairness perceptions; specifically, their findings suggest that fairness perceptions are lower for people with lower computer literacy.
\citet{pierson2017demographics}, along the same lines, found that students' views on algorithmic fairness changed by increasing algorithmic literacy through lecture and discussion: students ``became more likely to emphasize transparency, [and] more open to using algorithms rather than using judges.'' \cite{pierson2017demographics}
Finally, intuition tells us that AI-literate people may ``extract'' more information and understanding out of ADS explanations (e.g., because they know how supervised ML in general works).

\begin{itemize}
    \item[\textbf{H4}] People with higher AI literacy perceive an automated decision system to be more informationally fair than people with little or no knowledge in the field.
    \item[\textbf{H5}] People with higher AI literacy perceive an automated decision system to be more trustworthy than people with little or no knowledge in the field.
\end{itemize}

\section{Methodology}
We examine our hypotheses in the context of algorithmic lending.
We argue that this is a common context that affects many people at some point in life.
It is, furthermore, an area where ADS are typically already utilized within productive settings \cite{Atico2021,Infosys2019}.
Specifically, we confront study participants (SPs) with situations where a person was denied a loan.
Similar to \cite{binns2018s}, we argue that, in practice, explanations are much more likely to be requested by decision-subjects in response to negative outcomes; or, in other words: if someone gets the loan, interest in how and why exactly the decision was arrived at will likely drop.
However, we do by no means imply that reactions to positive outcomes are unworthy of being examined---given budget constraints, we defer them to future work.

\subsection{Study design}\label{sec:study_design}
We choose a between-subject design with the following conditions: first, we reveal to SPs some basic information about the lending company.
We then explain that a given individual’s loan application was rejected by the company, as well as that this decision was communicated to the applying individual electronically and in a timely fashion (see Fig.~\ref{fig:introductory_text} for the exact wording in our questionnaires).
Afterwards, we provide one of four explanations (i.e., conditions) to each SP.
Eventually, we measure the effects of assigning different conditions---and by design of the conditions, different amounts of information (AMTIN)---on two dependent variables: perceived informational fairness (INFF) and perceived trustworthiness (TRST) regarding the ADS.
(Recall that informational fairness perceptions do \emph{not} involve an actual assessment of the system's fairness w.r.t. its processes or outcomes.)
Additionally, we measure the (self-assessed) AI literacy (AILIT) of SPs.
We analyze whether differences in SPs’ AI literacy affect their perceptions.
All measurement items are summarized in \S~\ref{appendix:constructs_items}.
Note that for each construct, we measure multiple items; mostly drawn (and partially adapted) from prior work.

\paragraph{ADS Setup}
The ADS for our study consists of a random forest classifier which predicts loan approval on unseen data and is able to output different explanations.
For training our model, we utilize a publicly available dataset on home loan application decisions \cite{dataset2019}, which has been used in multiple data science competitions on \texttt{Kaggle}.
Note that comparable data---reflecting a given finance company’s individual circumstances and approval criteria---might in practice be used to train ADS \cite{Infosys2019}.
The dataset at hand consists of 614 labeled (loan Y/N) observations and includes the following features: \emph{applicant income, co-applicant income, credit history, dependents, education, gender, loan amount, loan amount term, marital status, property area, self-employment}.
After removing data points with missing values, 480 observations remain, 332 of which (69.2\%) involve the positive label (Y) and 148 (30.8\%) the negative label (N).
We used 70\% of the dataset to train our ADS and use the remaining 30\% as a holdout set for the experiment.
After encoding and scaling the features, we trained a random forest classifier with bootstrapping \cite{breiman2001random}, which achieves an out-of-bag accuracy estimate of 80.1\% on the held-out data.
We use this classifier's predictions on the holdout set as a basis for the upcoming conditions/explanations that the SPs are confronted with.
Since we are \emph{not} asking to assess the actual (procedural or distributive) fairness of the ADS, it is not critical to quantify how fair the system really is---any such effort would be highly contestable anyhow, for reasons of incompatible fairness notions \cite{chouldechova2017fair,kleinberg2016inherent,mulligan2019thing}.
The authors still (informally but independently) checked training data as well as output quality for any salient problems that may bias SPs' responses w.r.t. the dependent variables.

\begin{figure}[t]
    \centering
    \fcolorbox{black}{light-gray}{%
        \parbox{0.8\columnwidth}{%
            A finance company offers loans on real estate in urban, semi-urban, and rural areas. A potential customer first applies online for a specific loan, and afterwards, the company assesses the customer's eligibility for that loan.
            
            An individual applied online for a loan at this company. The company denied the loan application. The decision to deny the loan was communicated to the applying individual electronically and in a timely fashion.
        }%
    }
    \caption{Introduction of use case in questionnaires.}
    \label{fig:introductory_text}
\end{figure}

\paragraph{Explanations}
We impose several requirements on the explanations that we provide to SPs: overall, we employ only model-agnostic explanations \cite{adadi2018peeking} in a way that they could plausibly be provided to loan applicants (i.e., lay people) in real-world scenarios.
While explanations can be communicated in a wide variety of ways (see, e.g., \cite{adadi2018peeking,miller2019explanation,arrieta2020explainable,guidotti2018survey}), we confine ourselves to textual explanations (esp. no visuals) to control for differences in conveyance.
We also pick explanations that are immediately understandable semantically---this is important so as to collect meaningful responses.
On a related note, we ensure that explanations are not too long, in order to account for known issues around information overload \cite{bawden2009dark}.
Finally, and similar to \cite{binns2018s}, we pick explanations that can plausibly provide insights about a system's ``logic involved,'' as required, e.g., by the GDPR.
Based on these preliminaries, we assign SPs to one of four conditions that involve combinations of explanations w.r.t. (i) factors considered by the ADS, (ii) relative importance of these factors, and (iii) counterfactual scenarios where a rejected applicant would have been granted the loan.
We acknowledge that additional explanation styles would be equally interesting to consider; however, in order to keep the experiment size manageable, we must defer them to future work.

Our first condition, \emph{(Base)}, only reveals to the SPs that the loan decision was communicated to the applying individual electronically and in a timely fashion (as in Fig. \ref{fig:introductory_text}).
Apart from the \emph{(Base)} condition---which might be regarded as a black-box system---all other conditions include the additional information that the loan decision was made by an ADS (i.e., automated).
The second condition, \emph{(F)}, consists of disclosing the factors, including corresponding values for an observation (i.e., an applicant) from the holdout set whom our model denied the loan.
We refer to such an observation as a \emph{setting}.
In our study, we employ two different settings in each questionnaire, where settings are chosen at random from the pool of rejected applicants.
The authors, again, checked informally that no highly unusual (e.g., extreme outliers) settings were displayed that might distract SPs' perceptions and bias recorded responses.
Please refer to \S~\ref{appendix:explanation_styles} for an exemplary setting (introduction of use case plus conditions).
Next, we computed permutation feature importance \cite{breiman2001random} from our model and obtained the following hierarchy, using ``$\succ$'' as a shorthand for ``is more important than'': \emph{credit history $\succ$ loan amount $\succ$ applicant income $\succ$ co-applicant income $\succ$ property area $\succ$ marital status $\succ$ dependents $\succ$ education $\succ$ loan amount term $\succ$ self-employment $\succ$ gender}.
Revealing this ordered list in conjunction with \emph{(F)} makes up our third condition, \emph{(FFI)}.
To construct our fourth condition, we conducted an online survey with 20 quantitative and qualitative researchers to ascertain which of the aforementioned factors are actionable---in a sense that people can (hypothetically) act on them in order to increase their chances of being granted a loan.
According to this survey, the top-5 actionable factors are \emph{loan amount, loan amount term, property area, applicant income, co-applicant income}.
Our fourth condition \emph{(FFICF)} is then---in conjunction with \emph{(F)} and \emph{(FFI)}---the provision of three counterfactual scenarios where one actionable factor each is (minimally) altered such that our model predicts a loan approval instead of a rejection.
Our four conditions are summarized as follows:

\begin{table}[h!]
\centering
\begin{tabular}{p{1cm} p{6cm}}
\textit{(Base)} & Baseline without further explanations. \\
\textit{(F)} & Disclosure of factors. \\
\textit{(FFI)} & Disclosure of factors and factor importance. \\
\textit{(FFICF)} & Disclosure of factors, factor importance, and counterfactual scenarios.
\end{tabular}
\end{table}

Note that the order of provided explanations ($(Base)\rightarrow(F)\rightarrow(FFI)\rightarrow(FFICF)$) is not arbitrary: each subsequent condition provides the exact same information as the previous one \emph{and more}.
Since, e.g., factor importances implicitly reveal which factors the ADS considers, this would not necessarily hold true for, e.g., $(FI)\rightarrow(FIF)$.

\subsection{Data collection}

\begin{figure*}[t]
\centering
\begin{minipage}{0.24\textwidth}
    \centering
    \includegraphics[width=1\linewidth]{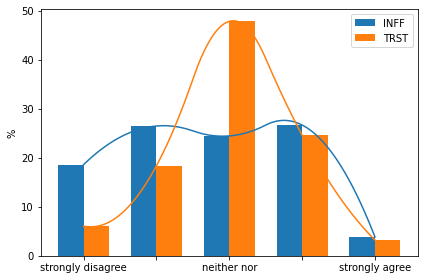}
    \caption*{\textit{(Base)}}
\end{minipage}%
\begin{minipage}{0.24\textwidth}
    \centering
    \includegraphics[width=1\linewidth]{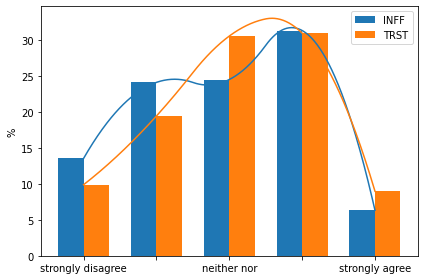}
    \caption*{\textit{(F)}}
\end{minipage}
\begin{minipage}{0.24\textwidth}
    \centering
    \includegraphics[width=1\linewidth]{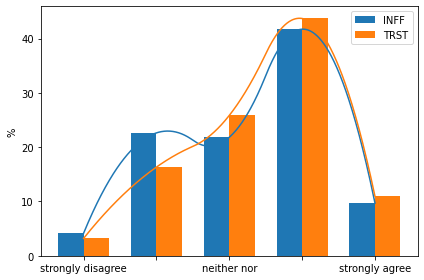}
    \caption*{\textit{(FFI)}}
\end{minipage}%
\begin{minipage}{0.24\textwidth}
    \centering
    \includegraphics[width=1\linewidth]{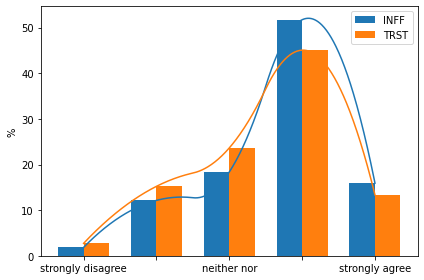}
    \caption*{\textit{(FFICF)}}
\end{minipage}%
\caption{Distributions of responses for informational fairness (INFF) and trustworthiness (TRST) per condition.}
\label{fig:distribution_responses}
\end{figure*}

Study participants (SPs) for our online study were (voluntarily) recruited via \texttt{Prolific} \cite{palan2018prolific} and asked to rate their agreement with multiple statements w.r.t. our dependent variables as well as their AI literacy on 5-point Likert scales---where 1 corresponds to ``strongly disagree'' and 5 denotes ``strongly agree''.
Additionally, we included multiple open-ended questions in the questionnaires to be able to better understand the reasoning behind SPs’ quantitative responses.
The SPs were randomly and in equal proportions assigned to one of the four conditions, and each SP was provided with two consecutive questionnaires associated with two different settings.
We collected 401 responses, of which 4 had to be eliminated due to failure to pass one or more attention checks.
Thus, we obtained 397 analyzable responses.
Among the SPs, 60\% indicated to be male, 39\% female, and the remaining SPs either responded with ``non-binary'' or chose not to disclose their gender; 46\% were students, 27\% employed full-time, 8\% employed part-time, 7\% self-employed, 11\% unemployed, less than 1\% retired, and 1\% chose not to disclose their profession.
The reported average age of SPs was 25.7.
SPs were monetarily compensated above the recommended min. pay of \$6.50 per hour.

\section{Quantitative Analyses and Results}\label{sec:quantitative_analysis}

We now examine the effects of our conditions and people's (self-assessed) AI literacy on perceived informational fairness and trustworthiness of our ADS.
For our measurement model, describing a confirmatory factor analysis and reporting correlations and factor loadings, we refer the reader to \S~\ref{app:measurement_model}.
In this section, we first present the results of group difference analyses for our conditions with tests for pairwise comparison.
After that, we report our findings on the validation of our hypotheses \textbf{H1} to \textbf{H5} with a full structural equation model.

\subsection{Analysis of group differences}\label{sec:group_differences}

Since we cannot confirm the assumption of normality for all variables, we conduct multiple non-parametric Kruskal-Wallis H tests for multiple group comparisons \cite{kruskal1952use}.
Afterwards, we carry out pairwise comparisons using Bonferroni-corrected Mann-Whitney U tests \cite{mann1947test}.
With these tests, we initially assess the effects of our four conditions revealing different amount of information (AMTIN) on the constructs of informational fairness (INFF) and trustworthiness (TRST).
Overall, we find a significant effect between different conditions on perceptions of informational fairness ($p < 0.001$) as well as on perceptions of trustworthiness ($p < 0.001$).
A Mann-Whitney U test for pairwise comparisons shows that the effect for informational fairness is significant ($p < 0.05$) between all conditions except \emph{(Base)} and \emph{(F)}.
The effect for trustworthiness is significant between \emph{(Base)} and \emph{(FFI)}, \emph{(Base)} and \emph{(FFICF)}, as well as \emph{(F)} and \emph{(FFICF)}, and marginally significant between \emph{(F)} and \emph{(FFI)} ($p = 0.052$).
Looking at the mean response values for (INFF) and (TRST) by condition (see Tab.~\ref{tab:means}), we note that they are increasing as more information is shown to SPs. Please refer to Fig.~\ref{fig:distribution_responses} for the distribution of responses by condition, and to Tab.~\ref{tab:mann_whitney} for a detailed summary of the results of the Mann-Whitney U tests.
\begin{table}[t]
\centering
\caption{Means and standard deviations of response values for informational fairness (INFF) and trustworthiness (TRST) by condition. All items were measured on 5-point Likert scales.}
\label{tab:means}
\begin{tabular}{c c c | c c}
\toprule
\multicolumn{1}{c}{\bf Condition} & \multicolumn{1}{c}{\bf M(INFF)} &
\multicolumn{1}{c}{\bf SD(INFF)} & \multicolumn{1}{c}{\bf M(TRST)} &
\multicolumn{1}{c}{\bf SD(TRST)}\\
\midrule
\textit{(Base)} & 2.71 & 1.16 & 3.01 & 0.89 \\
\textit{(F)} & 2.93 & 1.16 & 3.10 & 1.12 \\
\textit{(FFI)} & 3.30 & 1.05 & 3.43 & 0.99 \\
\textit{(FFICF)} & 3.68 & 0.94 & 3.51 & 0.99 \\
\midrule
\multicolumn{5}{l}{Notes: M = Mean; SD = Standard deviation} \\
\bottomrule
\end{tabular}
\end{table}

\begin{table*}[htbp]
\centering
\caption{Pairwise differences in perceptions of informational fairness (INFF) and trustworthiness (TRST) between conditions.}
\label{tab:mann_whitney}
\begin{tabular}{c c c | c c c}
\toprule
\multicolumn{3}{c}{\bf INFF} & \multicolumn{3}{c}{\bf TRST} \\
\midrule
\bf Condition 1 & \bf Condition 2 & \bf Difference & \bf Condition 1 & \bf Condition 2 & \bf Difference \\
\it(Base) & \it(F) & n/s & \it(Base) & \it(F) & n/s \\
\it(Base) & \it(FFI) & *** & \it(Base) & \it(FFI) & *** \\
\it(Base) & \it(FFICF) & *** & \it(Base) & \it(FFICF) & *** \\
\it(F) & \it(FFI) & * & \it(F) & \it(FFI) & n/s \\
\it(F) & \it(FFICF) & *** & \it(F) & \it(FFICF) & ** \\
\it(FFI) & \it(FFICF) & ** & \it(FFI) & \it(FFICF) & n/s \\
\midrule
\multicolumn{6}{l}{Notes: *$p<0.05$; **$p<0.01$; ***$p<0.001$; n/s: not significant} \\
\bottomrule
\end{tabular}
\end{table*}

\subsection{Hypotheses testing}
\begin{figure*}[htbp]
\centering
\includegraphics[width=0.7\linewidth,trim={2cm 2cm 2cm 3cm},clip]{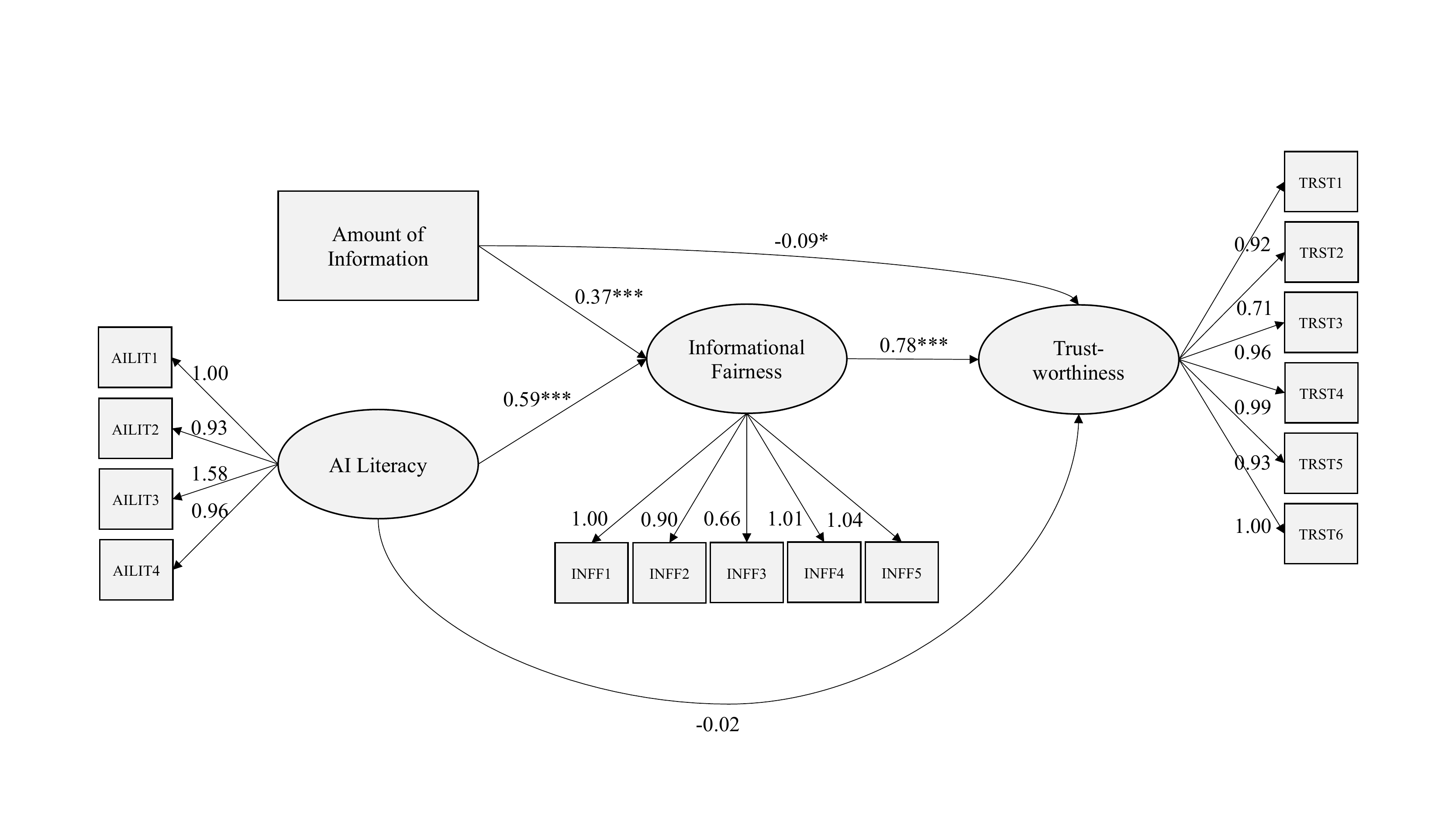}
\caption{Full structural equation model (SEM) including measurement model; *$p < 0.05$, **$p < 0.01$, ***$p < 0.001$.}
\label{fig:sem}
\end{figure*}

We estimate a full structural equation model (SEM), the results of which are depicted in Fig.~\ref{fig:sem}.
We also report more exhaustive information, including standard errors, z-values, p-values, and standardized path estimates in Tab.~\ref{tab:model_estimation} in \S~\ref{app:sem_detailed}.
Consistent with using Kruskal-Wallis H tests for group comparisons, we estimate our SEM using unweighted least squares (ULS) because this estimator makes no distributional assumptions.
We assess the fit of our model with multiple common measures: the comparative fit index (CFI) as well as Tucker-Lewis index (TLI) should be above 0.9 \cite{kline2015principles}, root mean square error of approximation (RMSEA) below 0.05 \cite{browne1992alternative}, and standardized root mean squared residual (SRMR) below 0.08 \cite{hair2016primer} to indicate good model fit. Our model's values are
\[
CFI = 0.997;\ TLI = 0.997;\ RMSEA = 0.024;\ SRMR = 0.051.
\]
Hence, all considered fit measures meet the required thresholds.
Note that the chi-square test is not a meaningful measure of model fit in our case because variables are not normally distributed, and because we apply the ULS method to estimate our model \cite{kenny2015measuring}.

In the following, we use a shorthand for our variables: AMTIN, AILIT, INFF, TRST (as introduced in \S~\ref{sec:study_design} and summarized in Tab.~\ref{tab:abbrev} of \S~\ref{sec:abbrev}).
To investigate our hypotheses, we first examine the effect of AMTIN on INFF.
As expected, and previously supported by the Kruskal-Wallis H test as well as the comparison of means between different conditions, increasing AMTIN has a significant positive effect on INFF (0.37***).
Hence, \textbf{H1} is supported.

Next, we examine the influence of AMTIN on TRST.
The results of the Kruskal-Wallis H test from \S~\ref{sec:group_differences} indicate that there is a significant positive relationship between AMTIN and TRST.
However, a mediation analysis within the SEM reveals that this effect is mediated by INFF.
When assessing this mediating effect more closely in the context of our SEM, a small direct effect of AMTIN on TRST persists.
Interestingly, in the context of the model, the stronger effect of AMTIN on TRST through INFF is positive, while the smaller but significant remaining direct effect is negative (-0.09*).
We discuss this in more detail in \S~\ref{sec:discussion}.
Overall, \textbf{H2}, which conjectures a positive \emph{total} (i.e., direct plus indirect) effect of AMTIN on TRST, is supported in our study.

The SEM’s path coefficient concerning \textbf{H3} (0.78***) confirms that there is a statistically significant positive relationship between INFF and TRST---which confirms \textbf{H3}.
This result provides a crucial individual piece of information in the context of the analysis of INFF as a mediator between AMTIN and TRST.
As presumed in \textbf{H4}, the path coefficient between AILIT and INFF (0.59***) confirms the conjecture of a significant positive relationship between these two variables---therefore, \textbf{H4} is supported by our results.
Similar to our findings w.r.t. the effect of AMTIN on TRST, the relationship between AILIT and TRST is also mediated by INFF.
The analysis of effects within the full SEM confirms a strong indirect effect of AILIT on TRST through INFF, but the remaining direct effect of AILIT on TRST is not significant.
Hence, the effect of AILIT on TRST is completely mediated by INFF.
In conclusion, \textbf{H5}, which assumes a positive relationship between AILIT and TRST, is supported.

\section{Qualitative Analysis}

In this section, we aim to understand people's perceptions in more detail.
To that end, we collected responses to open-ended questions regarding (i) what information SPs think they are missing (if any) to be able to judge whether the system behaves fairly, and (ii) SPs' perceptions of (in)appropriateness of the given explanations.
These questions were part of each condition.
The first and second author jointly coded the qualitative data according to grounded theory \cite{charmaz2003grounded}, i.e., codes evolved as we analyzed the data.
In total, 982 text passages were coded over five coding sessions with \texttt{MAXQDA} \cite{kuckartz2019analyzing}.
The emerging themes from the collected responses are summarized in the following subsections.
Every direct quote is provided with a unique identifier, introduced with the ``\#'' symbol.
Some responses contain statements w.r.t. multiple themes; hence, percentages do not always add up to 100\%.

\subsection{What information is missing?}\label{sec:info_missing}
\begin{figure}[htbp]
    \centering
    \includegraphics[width=0.8\linewidth]{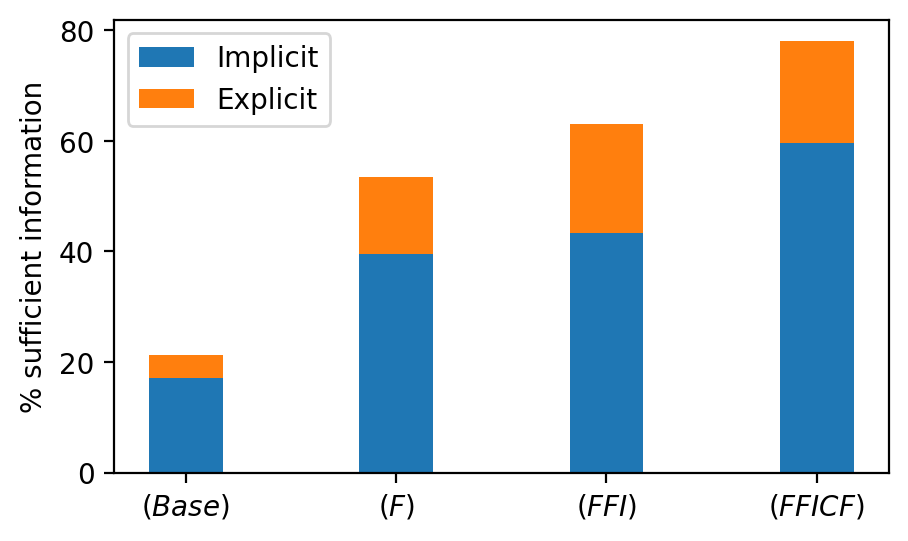}
    \caption{Percentage of responses indicating that study participants received sufficient information to judge whether the system's procedures are fair or unfair; either indicated explicitly in their responses, or implicitly by not answering the respective question.}
    \label{fig:sufficient_info}
\end{figure}

For this question, we coded 421 text passages from SPs' responses to the open-ended question: \emph{If you don't feel you received sufficient information to judge whether the decision-making procedures are fair or unfair, what information is missing?}
We distinguish responses by condition and examine how many SPs felt that they received sufficient information (either by saying so explicitly or by not answering this question altogether). The latter is visually summarized in Figure~\ref{fig:sufficient_info}.

\paragraph{(Base)}
Most SPs (79\%) assigned to this condition felt that they did not receive sufficient information; 17\% did not answer the question, and 4\% explicitly stated that they are not missing any information.
Little surprisingly, when asked which information they are missing, SPs were interested in knowing why the system made particular decisions; 37\% of all responses contained statements substantially similar to this: \emph{``All I know is that the loan was denied and not the reason why''} (\#1315).
Similarly, 30\% of responses inquired about decision criteria that underlie the rejected loans: \emph{``I have no way to know what references the company may or may not use to consolidate a decision about the eligibility of an individual for a particular loan, and therefore I might or might not find the procedures to be truly fair''} (\#1260).
16\% of responses also thought that decision-making procedures in general must be explained more thoroughly, arguing that \emph{``everything to do with how they made their decision of whether to accept the loan or not [is missing]''} (\#1234).
Some SPs were more specific as to what explanations they need: 18\% indicated that relevant factors of applicants would be helpful to know (\#1259: \emph{``To decide whether the decision-making procedures are fair or unfair, I probably would need to know how the client was economically and other factors such as criminal records''}); and 6\% of responses requested counterfactual-type insights related to recourse, e.g., \emph{``what he can do to try again''} (\#1265).

\paragraph{(F)}
In the \emph{(Factors)} condition, already 54\% of SPs indicated that they received sufficient information.
Of those who indicated that more information is needed, 15\% are still interested in the ``why'' behind the rejections (\#587: \emph{``I think clearly spelled reason is missing instead of numbers''}). 15\% still thought that more information w.r.t. decision criteria is needed.
Interestingly, knowing what factors are used by the ADS raises further, more specific, questions as to why (i) these given factors are considered (\#731: \emph{``There needs to be more in depth explanations given as to why these factors are taken into consideration''}), and (ii) not others, e.g., \emph{``how many loans have they taken out in the past, what is the money going to be spent on etc''} (\#663).
Overall, 23\% of SPs requested these justifications.
Another 10\% of responses indicated that it would be necessary to know how each factor impacts the final decision---both in terms of weighting (\#474: \emph{``What kind of value does each factor hold?''}) and monotonic relationships with the outcome (\#602: \emph{``The factors are told, but not which ones influenced the response positively of negatively.''})
Finally, 3\% are interested in counterfactual explanations, e.g., \emph{``how the factors should differ for the application to be approved''} (\#474).

\paragraph{(FFI)}
In this condition, only 37\% of SPs requested further information.
Among these, 15\% still requested more information w.r.t. reasons why the ADS rejected the applications; and 17\% felt that they still had not received sufficient information regarding decision-criteria (\#677: \emph{``There is not enough information about what thresholds have to be met to qualify for a loan.''})
On a related note, 6\% of SPs wanted to see more explanation as to why \emph{``the [factor importance] ranking is the way it is''} (\#764).
Similar to the \emph{(F)} condition, some SPs (10\%) wanted to know why certain factors of the applicants are not being considered by the ADS.
3\% of SPs still needed to know how exactly specific factors impact the final decision (\#684: \emph{`` I don't know the significance level/weight assigned to [the factors]''}); and another 3\% specifically requested counterfactual-type explanations.
A newly occurring theme is w.r.t. communication of the explanations, as 3\% requested \emph{``less formal descriptions''} (\#714).

\paragraph{(FFICF)}
In our condition with the highest amount of provided information, only 22\% requested additional information.
Generally speaking, responses are more dispersed compared to other conditions.
Some SPs still alluded to missing justification w.r.t. the given selection and importance of relevant factors (overall 14\%), and others (7\%) still asked for more information on the relationship between certain input factors and the outcome (\#796: \emph{``Since I think gender being a factor is unfair, not knowing the degree to which it affects the outcome seems to be a deficiency.''})
6\% of SPs were interested in the rationale behind providing given counterfactuals: \emph{``The factors that could have changed the outcome [are revealed], but not the reason why those [\dots] factors would be needed. Ex.: Why would a rural area be more easily accepted?''} (\#856)
Interestingly, no SP requested additional information as to why the ADS rejected the applicants---as opposed to the other conditions.
Yet, 11\% still requested more information w.r.t. decision criteria, e.g., \emph{``the thresholds that are required for a loan to be accepted''} (\#800).
6\% stated that processes were generally still not fully clear; however, some acknowledged that this might not necessarily be expedient, to begin with (\#863: \emph{``It's not clear how practically the priority system works, but I can understand it would be too hard to explain, and probably most of the people wouldn't understand it anyway.''})

\begin{figure}[htbp]
    \centering
    \includegraphics[width=1\linewidth,trim={0cm 0.3cm 0cm 0cm},clip]{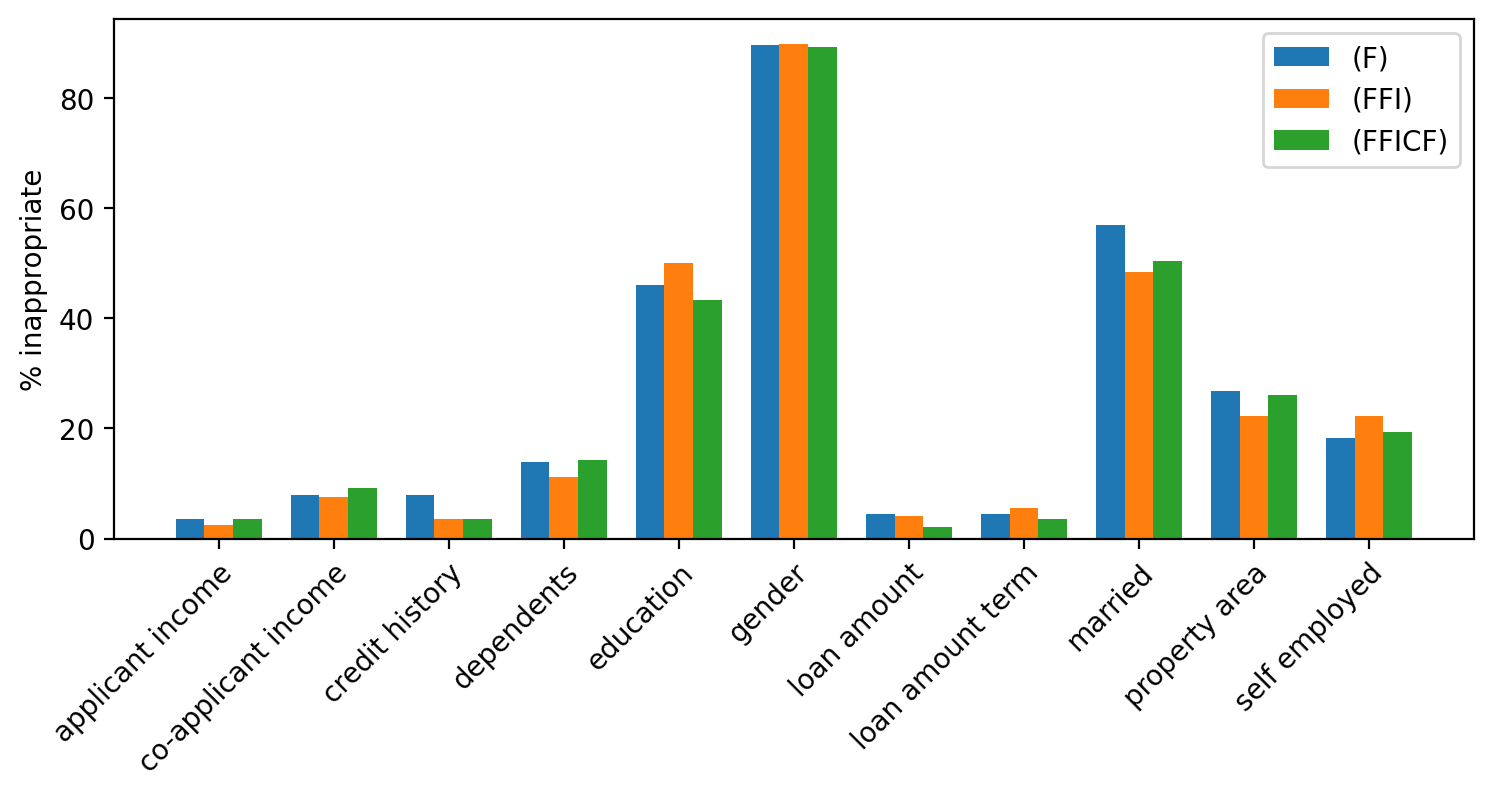}
    \caption{Inappropriate factors according to responses from study participants, broken down by condition.}
    \label{fig:inappropriate_features}
\end{figure}

\subsection{(In)Appropriateness of individual explanations}
We also asked SPs about their feelings of (in)appropriateness of isolated explanations, specific to the condition they were assigned to: \emph{Why do you think \{some factors, the order of factor importance, some counterfactual scenarios\} are appropriate or inappropriate?}
For that, we coded 561 text passages and summarized the main themes for each type of explanation.

\paragraph{Factors}%
Only 14\% of responses explicitly stated that (at least a subset of) the factors considered by the ADS were appropriate---mostly those related to an applicant's financial situation (\#602: \emph{``Economic factors seem apropriate [sic] to me. Self employment sometimes involves risks and it is a relevant factor also.''})
We also asked SPs to check specific factors they deem inappropriate---this is visualized (by condition) in Figure~\ref{fig:inappropriate_features}.
Among responses w.r.t. inappropriate factors, two general themes emerged: 72\% indicated that some factors are (causally) irrelevant for deciding on creditworthiness (\#632: \emph{``Some of the more social-oriented factors (ie education, gender, dependents) aren't necessarily indicative of someone's ability to pay back a loan''}), and 28\% found the usage of certain factors (primarily \emph{gender, education}, and \emph{married}) morally wrong (\#561: \emph{``In the world we live, i dont [sic] think gender is something to even be at question, neither marriage.''})
Interestingly, SPs often assumed that the sheer presence of a factor like \emph{gender} means that it is being used with malicious intent: \emph{``Gender can be somewhat problematic because all people deserve to have the right to the loan and not only men''} (\#637), or, \emph{``some factors like gender are plain racist to make a financial decision''} (\#647).

\paragraph{Factor importance}
Generally speaking, most SPs found the order of factor importance reasonably appropriate.
Many responses resembled this: \emph{``I may not agree with the placement of every single factor, but overall i think they are ranked appropriately''} (\#695).
Yet, 35\% still suggested concrete changes w.r.t. the order of importance; particularly around assigning less weight to \emph{education} and \emph{marital status}.
14\% were still entirely put off by the fact that \emph{gender} or \emph{marital status} were used in the decision-making process.
However, learning that \emph{gender} is the least important factor made many SPs feel better w.r.t. appropriateness of procedures (\#510: \emph{``It is appropriate. Gender should be considered the least and credit history is most important.''})
One SP even suggested that \emph{``gender could play a part in the decision making, but not a big one so it's good as it is''} (\#751). (Recall that \emph{gender} was ranked last in our explanation (see \S~\ref{sec:study_design}).)

\paragraph{Counterfactual scenarios}
47\% of coded responses indicated that the provided counterfactual scenarios are appropriate, e.g., endorsing that they \emph{``are all financial and based on the ability of the loan to be paid back''} (\#448).
However, 20\% questioned the effectiveness of adhering to some of the counterfactual recommendations; especially regarding suggested changes to \emph{co-applicant income} or \emph{property area}: \emph{``These factors do not change the fact that an applicant can or can not pay his/her debt''} (\#454).
\emph{Actionability} of counterfactual scenarios was another important theme: 9\% overall addressed this, being appreciative that some counterfactual scenarios are explicitly actionable (\#836: \emph{``Changing the loan term is possible immediately''}) and disenchanted when not (\#462: \emph{``Some hardly achivable [sic] scenarios must be met to ensure the bank [will] be repayed [sic].''})
Some themes were addressed by fewer SPs but are highly interesting: one SP was, e.g., confused by the ``direction'' of suggested changes: \emph{``Instead of a short loan amount term, it could be a bit longer''} (\#778).
Others were seemingly distracted by suggested changes that are (too) small: \emph{``The incomes are so close to the required that it shouldn't matter''} (\#447).
Finally, some SPs hinted at potential inconsistencies between individual explanations: \emph{``It seems odd that loan amount term is placed so low when it was one of the areas the individual could change to obtain the loan''} (\#435).

\section{Discussion and Implications}\label{sec:discussion}

In this section, we link our quantitative results to qualitative insights to get a better understanding as to why certain effects were observed, and we analyze and discuss in more detail the findings from the fitted SEM.
Finally, we allude to several implications of our work.

\paragraph{Connecting quantitative and qualitative findings}
As observed in Tab.~\ref{tab:means} (\S~\ref{sec:quantitative_analysis}), both perceptions of informational fairness and trustworthiness increase as more explanations are provided to SPs---however, INFF at a much higher rate than TRST.
Interestingly, many SPs in the \emph{(Base)} condition, who do not receive any further explanations w.r.t. the inner workings of the ADS, do \emph{not} find this ``black-box'' system to be overly problematic w.r.t. informational fairness: as can be seen in Fig.~\ref{fig:distribution_responses} (\S~\ref{sec:quantitative_analysis}), SPs' responses for INFF are approx. equally distributed across ratings 1--4.
This might be due to people's expectations; one SP simply stated that this \emph{``seems to be standard practice''} (\#1212) in terms of explaining ADS.
From Tab.~\ref{tab:mann_whitney} (\S~\ref{sec:quantitative_analysis}) we infer that providing relevant factors \emph{(F)} to SPs does not significantly increase INFF.
A likely reason for this observation is that SPs asked for significant follow-up information w.r.t. \emph{how} the factors are used for decision-making.
Both the differences for $(F)\rightarrow(FFI)$ and $(FFI)\rightarrow(FFICF)$ are significant for INFF.
Considering the qualitative findings (\S~\ref{sec:info_missing}), this seems little surprising as the complementary explanations (e.g., factor importance in \emph{(FFI)} over \emph{(F)}) were specifically requested by SPs.

While some explanations clearly helped SPs understand the given ADS better, they also reveal certain aspects that might be detrimental to people's trust.
Similar to INFF, one might have expected to see lower ratings for TRST in the \emph{(Base)} condition.
Instead, SPs' responses for TRST are symmetrically distributed around the mean of 3 (see Fig.~\ref{fig:distribution_responses}, \S~\ref{sec:quantitative_analysis}).
Regarding marginal effects of explanations on TRST, we note that none of $(Base)\rightarrow(F)$, $(F)\rightarrow(FFI)$, or $(FFI)\rightarrow(FFICF)$ lead to statistically significant changes in SPs' perceptions.
As for $(Base)\rightarrow(F)$, SPs' trust appears to be hampered by the experience that certain (presumably) inappropriate factors (e.g., \emph{gender}) are being considered by the ADS.
While the change $(F)\rightarrow(FFI)$ is marginally significant ($p = 0.052$) for TRST, we still suspect a certain attenuation due to SPs' disagreement with the relative importance ranking of certain factors like \emph{education} and \emph{married}.
On the other hand, from analyzing the qualitative statements, we might assume \emph{gender} playing the least important role in the decision-making process had a positive effect on SPs' trust.
As for $(FFI)\rightarrow(FFICF)$, we suspect that a potential positive effect of counterfactual explanations on perceived outcome control~\cite{houlden1978preference} might have been overshadowed by the fact that several SPs found some of the provided scenarios incomprehensible, ineffective, or unactionable.

\paragraph{Interpreting SEM results}
In addition to confirming significant \emph{total} effects (see Fig.~\ref{fig:sem}, \S~\ref{sec:quantitative_analysis}) of the amount of information (AMTIN) on INFF ($0.37$***) and TRST ($0.37\cdot0.78-0.09=0.20$***), we also learn that SPs' (self-assessed) AI literacy (AILIT) is strongly related to INFF ($0.59$***) and TRST ($0.44$***), implying that we observe higher INFF and TRST ratings for higher AI-literacy people---given our study setup.
Additionally, we see a strong positive relationship between INFF and TRST ($0.78$***).
The SEM also lets us decompose total effects of AMTIN and AILIT on TRST into direct and indirect (through the mediator INFF) effects (see Tab.~\ref{tab:effects}, \S~\ref{app:sem_detailed}).
We see, e.g., that the direct effect of AILIT on TRST ($-0.02$) is not significantly different from zero when INFF is acting as a mediator.
Since the indirect effect AILIT$\rightarrow$INFF$\rightarrow$TRST is significantly positive ($0.46$***), we observe a complete mediation of the effect of AILIT on TRST through INFF.
A similar observation can be made for the effect of AMTIN on TRST: the total effect consists of a significantly \emph{positive} indirect effect through INFF ($0.29$***) as well as a small \emph{negative} direct effect ($-0.09$*).
Hence, we conclude that increasing AMTIN does \emph{not directly} increase TRST, but that the positive total effect stems from the strong indirect effect through INFF.
This phenomenon is sometimes also referred to as \emph{inconsistent mediation}~\cite{kenny2015measuring,mackinnon2007mediation}. Future work should further investigate the link between INFF and TRST for other scenarios.

\paragraph{Implications}
Our work has several implications for the design of automated decision systems and explanations thereof.
Revealing to (potential) decision-subjects \emph{what} information about them is used and \emph{how} exactly individual factors affect the outcome is something that appears to go a long way towards facilitating informational fairness.
We have also seen that many people require an understanding of (assumed) monotonic relationships between individual features and outcome (\#856: \emph{``We don't know if being married is a good or bad thing in this case.''})
However, these types of global monotonic relationships cannot generally be derived from nonlinear ML models---something that has been discussed, e.g., in~\cite{rudin2019stop,schoeffer2021ranking,wang2020deontological}.
Employing inherently interpretable (e.g., linear) models might be a potential remedy.

We made a similar observation w.r.t. monotonicity for counterfactual explanations: people are put off when the ``direction'' of suggested change(s) contradicts commonly-held assumptions (e.g., if a \emph{decrease} in income were suggested in order to get the loan).
System designers must therefore pay close attention that counterfactual scenarios or general recommendations on recourse are intuitive, meaningful, and actionable.
Regarding the latter, we have observed that certain factors are deemed actionable by some SPs and immutable by others.
This poses further challenges w.r.t. individualizing explanations \cite{kuhl2020you}; this is also relevant for people with different AI backgrounds as their perceptions differ.
In general, however, counterfactual explanations appear to be effective in a way that they help people understand \emph{``where [an] applicant fell short''} (\#731).
From the analysis of qualitative data (also confirmed quantitatively), we learned that SPs in the \emph{(Base)} condition specifically requested explanations related to both factor importance and recourse / why the ADS decided negatively.
This suggests the employment of both explanation types in a complementary fashion.
Designers will have to ensure, however, that they are \emph{consistent} with one another.
For instance, people seem to expect that recommendations for recourse (e.g., that income should be increased) apply to the factors that are most important in the decision-making process.
Since individual explanations are often automatically and independently generated, this poses a significant technical challenge.
Our findings also suggest that informational fairness might be further increased by providing rejected loan applicants with a crisp statement in lay people's terms as to why they were denied.
Finally, regarding the usage of sensitive information like \emph{gender}, it should be clearly justified why and how (if at all) this information is used, and that this is not automatically to the disadvantage of marginalized groups; e.g., in the case of affirmative action~\cite{holzer2000assessing}.

\section{Limitations and Outlook}
We acknowledge limitations of our work that open up avenues for future studies.
Firstly, we investigated only one setting where ADS are currently used to inform consequential decisions: lending.
Our study design should be replicated and the results should be compared in different settings, e.g., hiring or university admissions, where the relevant factors will be significantly different.
It would also be interesting to work with domain experts, as opposed to crowdworkers.
Future work should further examine the complementarity and interplay of other explanation styles (e.g., case-based or demographic explanations~\cite{binns2018s}).
Furthermore, our quantitative results (including SEM) are contingent upon the concrete instantiation of our ADS including the employed explanations, which limits our ability to generalize findings.

While we informally checked the model as well as the underlying data and all derived explanations so as to ensure behavior that might be representative of many real-world applications, it would be insightful to randomize different aspects about the model's quality and compare the results.
More specifically, if we managed to construct---broadly speaking---a trustworthy ADS and an untrustworthy ADS, we would be able to contrast people's perceptions for either system.
This would allow to derive insights w.r.t. \emph{(un)warranted} perceptions, i.e., (i) are people \emph{actually} able to spot problematic behavior of ADS, and (ii) do they trust the system if and only if the system is trustworthy?
In fact, for an untrustworthy ADS, we would ideally expect that more explanations lead to \emph{higher} informational fairness perceptions but to \emph{lower} trust.
If perceptions of trustworthiness increase \emph{regardless} of the actual trustworthiness of the ADS, this would indicate serious issues around over-reliance \cite{Skitka2000} or automation bias \cite{Goddard2014,de2020case}, and must be avoided by system designers at all costs.

We also acknowledge that our work does not explicitly take into account potential issues around information overload~\cite{bawden2009dark}: while we specifically examine situations where selected explanations convey complementary information, unsystematic provision of more and more explanations will likely have undesirable effects.
The authors suggest by no means that more information is always better.
Finally, we hope that this work can serve as a stepping stone for further empirical research on the complementarity and interplay of different explanations and their effects on people's perceptions towards ADS.

\balance

\begin{acks}
We thank our study participants as well as our anonymous reviewers, who helped improve this manuscript.
\end{acks}

\bibliographystyle{ACM-Reference-Format}
\bibliography{bibliography}

\appendix
\onecolumn

\section{Abbreviations}\label{sec:abbrev}

Tab.~\ref{tab:abbrev} contains our most commonly used abbreviation.

\begin{table}[htb]
\centering
\caption{Summary of commonly used abbreviations.}
\label{tab:abbrev}
\begin{tabular}{l l}
\toprule
\bf Abbreviation & \bf Explanation \\
\midrule
ADS & Automated decision system(s) \\
AILIT & AI literacy \\
AMTIN & Amount of information \\
\textit{(Base)} & Baseline treatment without explanations \\
\textit{(F)} & Treatment with disclosure of factors \\
\textit{(FFI)} & Treatment with disclosure of factors and factor importance \\
\textit{(FFICF)} & Treatment with disclosure of factors, factor importance, and counterfactual explanations \\
INFF & Informational fairness (dependent variable) \\
SEM & Structural equation model \\
SP & Study participant(s) \\
TRST & Trustworthiness (dependent variable) \\
XAI & Explainable AI \\
\bottomrule
\end{tabular}
\end{table}

\section{Constructs and Measurement Items}\label{appendix:constructs_items}

All items within the following constructs were measured on a 5-point Likert scale and mostly drawn (and adapted) from previous studies. 

\begin{enumerate}
    
    \item \textbf{Informational Fairness (INFF)}
    
    \begin{itemize}
        \item The automated decision system explains decision-making procedures thoroughly. \citep{colquitt2015measuring}
        \item The automated decision system’s explanations regarding procedures are reasonable. \citep{colquitt2015measuring}
        \item The automated decision system tailors communications to meet the applying individual’s needs. \citep{colquitt2015measuring}
        \item I understand the process by which the decision was made. \citep{binns2018s}
        \item I received sufficient information to judge whether the decision-making procedures are fair or unfair.
    \end{itemize}
    
    \item \textbf{Trustworthiness (TRST)}
    
    \begin{itemize}
        \item Given the provided explanations, I trust that the automated decision system makes good-quality decisions. \citep{lee2018understanding}
        \item Based on my understanding of the decision-making procedures, I know the automated decision system is not opportunistic. \citep{chiu2009understanding}
        \item Based on my understanding of the decision-making procedures, I know the automated decision system is trustworthy. \citep{chiu2009understanding}
        \item I think I can trust the automated decision system. \citep{carter2005utilization}
        \item The automated decision system can be trusted to carry out the loan application decision faithfully. \citep{carter2005utilization}
        \item In my opinion, the automated decision system is trustworthy. \citep{carter2005utilization}
    \end{itemize}
    
    \item \textbf{(Self-Assessed) AI Literacy (AILIT)}
    
    \begin{itemize}
        \item How would you describe your knowledge in the field of artificial intelligence?
        \item Does your current employment include working with artificial intelligence?
        \item I am confident interacting with artificial intelligence. \citep{wilkinson2010construction}
        \item I understand what the term \emph{artificial intelligence} means.
    \end{itemize}
    
\end{enumerate}

\section{Explanation Styles for One Exemplary Setting}\label{appendix:explanation_styles}

\begin{tcolorbox}[breakable, enhanced jigsaw, sharp corners, opacityback=0, fonttitle=\bfseries, colframe=black, boxrule=1pt, title=Condition \textit{(F)}]

\begin{tcolorbox}[enhanced jigsaw, sharp corners, colframe=black, boxrule=1pt]
A finance company offers loans on real estate in urban, semi-urban and rural areas. A potential customer first applies online for a specific loan, and afterwards the company assesses the customer's eligibility for that loan.

An individual applied online for a loan at this company. The company denied the loan application. The decision to deny the loan was made by an automated decision system and communicated to the applying individual electronically and in a timely fashion.
\end{tcolorbox}

\tcblower

The automated decision system explains that the following factors (in alphabetical order) on the individual were taken into account when making the loan application decision:

\begin{itemize}
    \item Applicant Income: \$3,069 per month
    \item Co-Applicant Income: \$0 per month
    \item Credit History: Good
    \item Dependents: 0
    \item Education: Graduate
    \item Gender: Male
    \item Loan Amount: \$71,000
    \item Loan Amount Term: 480 months
    \item Married: No
    \item Property Area: Urban
    \item Self-Employed: No
\end{itemize}
\end{tcolorbox}

\begin{tcolorbox}[breakable, enhanced jigsaw, sharp corners, opacityback=0, fonttitle=\bfseries, colframe=black, boxrule=1pt, title=Condition \textit{(FFI)}]

\begin{tcolorbox}[enhanced jigsaw, sharp corners, colframe=black, boxrule=1pt]
A finance company offers loans on real estate in urban, semi-urban and rural areas. A potential customer first applies online for a specific loan, and afterwards the company assesses the customer's eligibility for that loan.

An individual applied online for a loan at this company. The company denied the loan application. The decision to deny the loan was made by an automated decision system and communicated to the applying individual electronically and in a timely fashion.
\end{tcolorbox}

\tcblower

The automated decision system explains \dots 
\begin{itemize}
    \item \dots that the following factors (in alphabetical order) on the individual were taken into account when making the loan application decision:

    \begin{itemize}
        \item Applicant Income: \$3,069 per month
        \item Co-Applicant Income: \$0 per month
        \item Credit History: Good
        \item Dependents: 0
        \item Education: Graduate
        \item Gender: Male
        \item Loan Amount: \$71,000
        \item Loan Amount Term: 480 months
        \item Married: No
        \item Property Area: Urban
        \item Self-Employed: No
    \end{itemize}
    \item \dots that different factors are of different importance in the decision. The following list shows the order of factor importance, from most important to least important:
    Credit History $\succ$ Loan Amount $\succ$ Applicant Income $\succ$ Co-Applicant Income $\succ$ Property Area $\succ$ Married $\succ$ Dependents $\succ$ Education $\succ$ Loan Amount Term $\succ$ Self-Employed $\succ$ Gender
\end{itemize}
\end{tcolorbox}

\begin{tcolorbox}[breakable, enhanced jigsaw, sharp corners, opacityback=0, fonttitle=\bfseries, colframe=black, boxrule=1pt, title=Condition \textit{(FFICF)}]

\begin{tcolorbox}[enhanced jigsaw, sharp corners, colframe=black, boxrule=1pt]
A finance company offers loans on real estate in urban, semi-urban and rural areas. A potential customer first applies online for a specific loan, and afterwards the company assesses the customer's eligibility for that loan.

An individual applied online for a loan at this company. The company denied the loan application. The decision to deny the loan was made by an automated decision system and communicated to the applying individual electronically and in a timely fashion.
\end{tcolorbox}

\tcblower

The automated decision system explains \dots 
\begin{itemize}
    \item \dots that the following factors (in alphabetical order) on the individual were taken into account when making the loan application decision:

    \begin{itemize}
        \item Applicant Income: \$3,069 per month
        \item Co-Applicant Income: \$0 per month
        \item Credit History: Good
        \item Dependents: 0
        \item Education: Graduate
        \item Gender: Male
        \item Loan Amount: \$71,000
        \item Loan Amount Term: 480 months
        \item Married: No
        \item Property Area: Urban
        \item Self-Employed: No
    \end{itemize}
    \item \dots that different factors are of different importance in the decision. The following list shows the order of factor importance, from most important to least important:
    Credit History $\succ$ Loan Amount $\succ$ Applicant Income $\succ$ Co-Applicant Income $\succ$ Property Area $\succ$ Married $\succ$ Dependents $\succ$ Education $\succ$ Loan Amount Term $\succ$ Self-Employed $\succ$ Gender
    \item \dots that the individual would have been granted the loan if---everything else unchanged---one of the following hypothetical scenarios had been true:
    \begin{itemize}
        \item The Co-Applicant Income had been at least \$800 per month
        \item The Loan Amount Term had been 408 months or less
        \item The Property Area had been Rural
    \end{itemize}
\end{itemize}
\end{tcolorbox}

\section{Measurement Model}\label{app:measurement_model}
In order to assess the validity and the reliability of our constructs, we conduct a confirmatory factor analysis and assess the results w.r.t. multiple measures.
As measures for convergent reliability, we examine average variance extracted (AVE) and composite reliability (CR).
For the constructs of informational fairness and trustworthiness, AVE is above the recommended threshold of 0.5, whereas the AVE of AI literacy is 0.41.
According to \citet{fornell1981evaluating}, if AVE is low, convergent validity of a construct can still be sufficient if composite reliability (CR) is above 0.6, which is the case for all three constructs, including AI literacy (see Tab.~\ref{tab:corr_meas}).
In fact, the CR of our three main constructs, informational fairness (0.88), trustworthiness (0.94), and AI literacy (0.72) is above the recommended threshold of 0.7 \cite{barclay1995partial}, indicating that our convergent validity is adequate for AI literacy as well, despite the lower AVE measure. 

Cronbach’s alpha (CA) values for our constructs are larger than the recommended threshold of 0.7, thus showing good reliability for all constructs \cite{cortina1993coefficient}.
Validity and reliability measures are summarized in Tab.~\ref{tab:corr_meas}.
Our matrix of factor loadings, demonstrated in Tab.~\ref{tab:loadings}, shows that all items load highly ($>$0.5) on one factor each with low cross-loadings, and the correlations between factors are all below 0.7 (see Tab.~\ref{tab:corr_meas}).
Furthermore, the AVE value of each of our constructs is larger than the squared correlation of that construct with every other construct, which is a discriminant validity measure suggested by \citet{chin1998partial} and \citet{fornell1981evaluating}.
Therefore, convergent validity and discriminant validity are sufficiently satisfied.
We test for multicollinearity by determining the variance inflation factors (VIF).
According to a rule of thumb, the VIF has to be lower than 10, otherwise, multicollinearity might be a serious problem \cite{vittinghoff2011regression}.
All VIFs in our model are less than 2, which indicates that there are no issues of multicollinearity.

\begin{table}[thb]
\centering
\caption{Correlations and measurement information for latent factors.}
\label{tab:corr_meas}
\begin{tabular}{c c c c c c c c c}
\toprule
\bf Factor & \bf M & \bf SD & \bf CA & \bf CR & \bf AVE & \bf INFF & \bf TRST & \bf AILIT \\
\midrule
INFF & 3.15 & 0.87 & 0.87 & 0.88 & 0.60 & 1.00 & & \\
TRST & 3.26 & 0.84 & 0.94 & 0.94 & 0.73 & 0.67 & 1.00 &  \\
AILIT & 2.87 & 0.61 & 0.71 & 0.72 & 0.41 & 0.25 & 0.18 & 1.00 \\
\midrule
\multicolumn{9}{l}{Notes: M = Mean; SD = Standard deviation} \\
\bottomrule
\end{tabular}
\end{table}

\begin{table}[htb]
\centering
\caption{Standardized loadings of measurement items on constructs.}
\label{tab:loadings}
\begin{tabular}{c c c c}
\toprule
\bf Measurement item & \bf INFF & \bf TRST & \bf AILIT \\
\midrule
INFF1 & \bf 0.95 & -0.11 & -0.03 \\
INFF2 & \bf 0.65 & 0.21 & 0.01 \\
INFF3 & \bf 0.52 & 0.10 & 0.05 \\
INFF4 & \bf 0.79 & 0.01 & 0.03 \\
INFF5 & \bf 0.76 & 0.01 & 0.00 \\
\midrule
TRST1 & 0.24 & \bf 0.66 & -0.05 \\
TRST2 & 0.20 & \bf 0.51 & -0.08 \\
TRST3 & 0.01 & \bf 0.90 & -0.01 \\
TRST4 & -0.08 & \bf 0.97 & 0.06 \\
TRST5 & 0.02 & \bf 0.90 & 0.05 \\
TRST6 & -0.09 & \bf 1.01 & 0.00 \\
\midrule
AILIT1 & 0.08 & -0.11 & \bf 0.73 \\
AILIT2 & 0.06 & -0.03 & \bf 0.53 \\
AILIT3 & -0.12 & 0.17 & \bf 0.67 \\
AILIT4 & 0.00 & -0.02 & \bf 0.58 \\
\bottomrule
\end{tabular}
\end{table}

\section{SEM Model: Results of Model Estimation}\label{app:sem_detailed}
Detailed information on the results of the SEM model estimation, including path estimates, standard errors (SE), z-values, p-values, and standardized estimates (Std.lv) are reported in Tab.~\ref{tab:model_estimation}.
A breakdown of direct and indirect effects of independent variables on trustworthiness (TRST) is given in Tab.~\ref{tab:effects}.

\begin{table}[thb]
\centering
\caption{Results of model estimation.}
\label{tab:model_estimation}
\begin{tabular}{c c c c c c}
\toprule
\bf Path & \bf Estimate & \bf SE & \bf z-value & \bf p-value & \bf Std.lv \\
\midrule
AILIT $\rightarrow$ INFF & 0.59*** & 0.08 & 7.01 & $<$0.001 & 0.31 \\
AMTIN $\rightarrow$ INFF & 0.37*** & 0.03 & 14.25 & $<$0.001 & 0.47 \\
INFF $\rightarrow$ TRST & 0.78*** & 0.05 & 15.30 & $<$0.001 & 0.78 \\
AILIT $\rightarrow$ TRST & -0.02 & 0.07 & -0.24 & 0.81 & -0.01 \\
AMTIN $\rightarrow$ TRST & -0.09* & 0.04 & -2.55 & 0.01 & -0.11 \\
\midrule
\multicolumn{6}{l}{Notes: *$p<0.05$; **$p<0.01$; ***$p<0.001$} \\
\bottomrule
\end{tabular}
\end{table}

\begin{table}[thb]
\centering
\caption{Decomposition of effects on perceived trustworthiness.}
\label{tab:effects}
\begin{tabular}{c c c c}
\toprule
& \bf Direct effect & \bf Indirect effect & \bf Total effect \\
\midrule
AMTIN on TRST & -0.09* & 0.37$\cdot$0.78=0.29*** & 0.20*** \\
AILIT on TRST & -0.02 & 0.59$\cdot$0.78=0.46*** & 0.44*** \\
\midrule
\multicolumn{4}{l}{Notes: *$p<0.05$; **$p<0.01$; ***$p<0.001$} \\
\bottomrule
\end{tabular}
\end{table}

\section{Software and Tools}
Tab.~\ref{tab:software_tools} contains all employed software and tools.

\begin{table}[thb]
\centering
\caption{Software and tools.}
\label{tab:software_tools}
\begin{tabular}{l l l}
\toprule
\bf Task(s) & \bf Software/tool & \bf Source \\
\midrule
Data processing (general) & \texttt{Python} & \citet{van1995python} \\
ML for training ADS and predictions & \texttt{Python} package \texttt{scikit-learn} & \citet{pedregosa2011scikit} \\
Crowdsourcing study participants & \texttt{Prolific} & \citet{palan2018prolific} \\
Questionnaires & \texttt{SoSci Survey} & \citet{leiner2019sosci} \\
Survey data processing, statistical analyses & \texttt{R} & \citet{coreteam2017R} \\
CFA, model fit, measurement model, SEM & \texttt{R} package \texttt{lavaan} & \citet{rosseel2012lavaan} \\
Fit measures, reliability measures & \texttt{R} package \texttt{cSEM} & \citet{rademaker2020cSEM} \\
Cross-loadings table, correlations & \texttt{R} package \texttt{psych} & \citet{revelle2020psych} \\
VIF & \texttt{R} package \texttt{car} & \citet{fox2019car} \\
Qualitative analysis & \texttt{MAXQDA} & \citet{kuckartz2019analyzing} \\
\bottomrule
\end{tabular}
\end{table}

\end{document}